\documentclass[manuscript]{aastex}

\slugcomment{To appear in Ap.J.}

\shorttitle{Polarization Diagnostics for Cluster Line-emission}
\shortauthors{Sparks et al.}

\begin{document}


\title{Polarization Diagnostics for Cool Core Cluster Emission Lines}


\author{W.B. Sparks\altaffilmark{1}, J.E. Pringle\altaffilmark{1,2}, R.F. Carswell\altaffilmark{2}, G.M. Voit\altaffilmark{3}, M. Donahue\altaffilmark{3}, M. Cracraft\altaffilmark{1}, E. Meyer\altaffilmark{1}, J.H. Hough\altaffilmark{4}, N. Manset\altaffilmark{5}}
\affil{1. Space Telescope Science Institute, 3700 San Martin Drive, Baltimore, MD 21218}
\affil{2. Institute of Astronomy, University of Cambridge, Madingley Road, Cambridge, CB3~0HA, UK}
\affil{3. Dept of Physics and Astronomy, Michigan State University, East Lansing, MI~48824-2320}
\affil{4. University of Hertfordshire, College Lane, Hatfield, AL10 9AB, UK.}
\affil{5. CFHT, 65-1238 Mamalahoa Hwy, Kamuela, HI 96743}
\email{sparks@stsci.edu}




\begin{abstract}
The nature of the interaction between low-excitation gas filaments at $\sim 10^4$K, seen in optical line emission, and diffuse X-ray emitting coronal gas at $\sim 10^7$K in the centers of galaxy clusters remains a puzzle.
The presence of a strong, empirical correlation between the two gas phases is indicative of a fundamental relationship between them, though as yet of undetermined cause.
The cooler filaments, originally thought to have condensed from the hot gas, could also arise from a merger or the disturbance of cool  circumnuclear gas by nuclear activity.
Here, we have searched for intrinsic line emission polarization in cool core galaxy clusters as a diagnostic of fundamental transport processes.
Drawing on developments in solar astrophysics, direct energetic particle impact induced polarization holds the promise to definitively determine the role of collisional processes such as thermal conduction in the ISM physics of galaxy clusters, while  providing insight into other highly anisotropic
excitation mechanisms such as shocks, intense radiation fields and suprathermal particles.
Under certain physical conditions, theoretical calculations predict of order ten percent polarization.
Our observations of the  filaments in four nearby cool core clusters place stringent upper limits ($\lesssim 0.1\%$) on the presence of emission line polarization, requiring that if thermal conduction is operative, the thermal gradients are not in the saturated regime. This limit is consistent with theoretical models of the thermal structure
of filament interfaces.

\end{abstract}





\section{Introduction}

The close correlation between cool cores viewed in X-rays and optical emission line nebulae in clusters of galaxies has been recognized for many years, but the physical reason for this connection remains unclear\footnote{Based on observations collected at the European Organisation for Astronomical Research in the Southern Hemisphere, Chile, programme 086.B-0138A}. Optical emission line filamentary structures have been seen and analyzed in many cool core clusters (e.g. Abell 2597: Voit \&\ Donahue 1997, Donahue et al. 2000; PKS0745-19 Donahue et al. 2000; NGC 4696: Sparks et al. 1989, Crawford et al. 2005, ; M87: Sparks et al. 1993, 2004; NGC 1275: Conselice, Gallagher \& Wyse 2001; Hatch et al. 2006; McDonald et al. 2010 and many others). While the origin of the optical emission line filaments and the mechanisms responsible for their ionization have been extensively studied, a consensus on the dominant physics has not emerged.

Many excitation mechanisms for the optical filaments have been considered, including photoionization by the central AGN, photoionization by hot stars, excitation by shocks, and energy transport from the hot coronal ISM in which they are embedded. Voit \&\ Donahue (1997) concluded that ``neither shocks nor photoionization alone can reproduce the emission line intensity ratios'' and that some additional source of heating must be present. Similar conclusions are reached by Hatch et al. (2007) for a number of cool core clusters, and in M87 we showed that hot stars were simply not present in the vicinity of the filaments (Sparks et al. 2009). Renewed interest in these areas has emerged with the possibility that feedback from the AGN into the hot coronal ISM is important, and hence it is critical to understand the transport processes in these environments and how apparently disparate gas phases are, in fact, related.

One obvious source of extra heating comes from the fact that the cool $H\alpha$ emitting gas is situated in a surrounding hot X-ray emitting medium. Thus thermal conductivity is a strong candidate (Sparks et al. 1989; Ferland et al. 2009). We recently discovered hot gas at $10^5$K associated with low temperature $H\alpha+[NII]$ optical filaments in M87 (Sparks et al. 2009), as predicted by models invoking thermal conduction (Nipoti \&\ Binney 2004). The concept of conduction is gaining in popularity (Ferland et al. 2009; Hudson et al. 2010; Werner et al. 2013). Hudson et al. (2010) offer as possible heating models ``conduction, central AGN heating via direct cosmic ray-ICM interaction+conduction, AGN heating by bubble induced weak shocks, soundwaves+conduction and turbulence+conduction''.

The current work is targeted towards determining whether thermal conduction is effective in cool core galaxy clusters by applying innovative techniques drawn from solar physics, to analogous scenarios on extragalactic scales.
In plasmas, if the excitation of the atoms that radiate line emission is a strongly anisotropic process, then the ensemble of atoms retains a memory of that directionality and the consequent line emission can be polarized. Strongly anisotropic excitation processes include powerful anisotropic radiation fields, shocks, and the energetic electrons and protons arising from steep temperature gradients associated with thermal conduction, particularly in the saturated regime. Different types of line emission such as resonance lines, permitted recombination lines, and collisionally excited forbidden lines, respond in different ways to these stimuli and the consequent line polarization contains potentially crucial information on the underlying physics (Landi Degl'Innocenti \&\ Landolfi 2004).

Polarization levels of up to 20-30\%\ have been claimed for Solar flare and prominence emission (Henoux et al. 1983a; L\'opez Ariste 2005; Firstova et al. 2008) and 5-30\%\ in auroral emissions of the Earth (Duncan 1959; Lilensten et al. 2008). In solar physics, the strong temperature gradients of thermal conduction result in a large collisional anisotropy and the optical polarization can be directly related to the heat flux relative to its saturated value. Saturated conduction gives a line polarization of 8\%, relatively easy to detect (Henoux et al. 1983b; Amboudarham et al. 1992; see Fig. 4). This physical process is  analogous to that which may dominate the physics of galaxy cluster cores, and directly relates the observable (polarization) to the theoretical stimulus (conduction, either saturated or unsaturated). Other physical mechanisms can cause emission-line polarization, including strongly anisotropic ionizing radiation, fast shocks (Laming 1990b predicts $H\alpha$ polarization up to 10\%\ in SNR) and non-thermal particles (Henoux et al. 2003; Ferland et al. 2009), each with their own distinctive character and polarization properties. Our aim is to obtain, for the first time, empirical measurements or constraints on the actual level of polarization in the optical emission filaments of cool core clusters and introduce an important new physical diagnostic that could revolutionize our understanding of the plasma physics of cool core galaxy clusters since it bears directly on the relevance of fundamental physical processes.

\section{OBSERVATIONS}

\subsection{Targets}

Our targets are well-studied central cluster galaxies with extensive low excitation filament systems, moderate power radio emission and clear indications of interaction between the X-ray and radio plasma. Observations utilize a $22\times 2$~arcsec spectrograph slit, configured as described in \S~2.2.

{\bf M87:} has a well-known low ionization optical filament system distributed around the periphery of the inner radio lobes and jet. We recently discovered FUV CIV
line emission exactly coincident with this material, but  arising from gas a factor of ten higher in temperature, consistent with our prediction for a model in
which the filaments are excited by Spitzer thermal conduction from the hot coronal gas (Sparks et al. 2009; Sparks et al. 2012). Spectropolarimetry observations were acquired with three slit locations and orientations, as shown in Fig.~1.

{\bf NGC 4696:} is the central dominant galaxy in the classical cool-core Centaurus Cluster. This well-studied system formed the basis of the suggestion by Sparks et al. (1989) that mergers in the presence of thermal conduction can be the cause of the X-ray excess and line emission, and not cooling flows. The optical filaments are dusty with normal extinction characteristics indicative of a merger origin. The X-ray morphology is similar to the optical though extended over a much larger scale, and the radio emission is compact and has a steep spectrum, on similar scale to the optical. Fig.~2 shows the two slit positions used.

{\bf PKS 0745-19:} Heckman et al. (1989), Donahue et al. (2000), Wilman et al. (2009) show that the optical line emission forms a roughly triangular or conical shape to the West of the nucleus, with two dominant arms of emission. The radio source is irregular and coincident in scale with the optical filaments, located primarily to the South. Slits centered on the nucleus with position angles 90$^{\circ}$ and 45$^{\circ}$ were used, Fig.~3.

{\bf Hydra A:} displays the now classical feedback situation (McNamara et al. 2000) with X-ray cavities unarguably at the locations of the outwardly propagating radio jets.
The optical line emitting gas spans the region between the radio core and radio knots, shares the S-symmetry of the radio emission and overlaps with the radio emission
only at the edges of the knots (Baum et al. 1988). Our long slit, in p.a.$\approx 20^{\circ}$, runs approximately in the direction of the radio source, orthogonal to an edge-on dust disk, Fig.~4.

The filaments in all of these targets span a range of brightnesses and there is a range of line-ratios within the filaments. Although we know that filaments are dusty, we do not expect significant polarization either from scattering or dichroic absorption. In M87 for example the optical depth $\tau\approx 0.01$ (Sparks, Ford \&\ Kinney 1993) which would result in a net dichroic polarization for Galactic dust of $p_{max}=0.09 E(B-V)$, hence $p< 0.03\%$. Even if dust characteristics differ from Galactic, we expect to recognize dust through its additional effect on the background galaxy continuum polarization and not just at wavelengths of optical line emission.

\subsection{Observations}

We obtained VLT UT1 spectropolarimetry of optical emission filaments using FORS2 in long slit spectropolarimetric mode. In this mode the light entering the spectrograph encounters a polarization slit mask, a rotatable half wave plate, a Wollaston prism to split the polarization o- and e-beams, and a grism to disperse the light.
The polarization mask, required so that the dual polarization beams do not overlap on the detector, results in a long slit comprised of 22 arcsec segments. For our analysis we used only the single 22~arcsec segment centered on the target of interest, as illustrated in Figs.~1---4. We used the 300V+10 grism to obtain spectra from $\approx 450$~nm to $\approx 900$~nm, and the GG435 order sorting filter. The slit width was 2~arcsec, chosen to be wide in order to maximize the light gathered on the detector.

For a given half wave plate setting we acquire a single Stokes parameter, and to reduce systematics, the same Stokes parameter is observed with the beams reversed using rotation of the half wave plate. Hence at least four wave plate rotations are required for a complete set of linear Stokes polarization spectra. Our observations used half wave plate rotation angles of 0, 22.5, 45, and 67.5 degrees.

The chosen spectral window encompasses the strong low excitation red emission lines, $H\alpha$, $[NII]$6548,6584, $[SII]$6717, 6730, and $[OI]$6300 and provides adequate spectral resolution to separate them, though the $H\alpha$ and $[NII]$ lines overlap. Weaker lines  include $[OIII]$, HeI, $[NI]$. The mix of recombination and collisional lines in principle allows us to contrast any polarization found between the different excitation mechanisms. All observations were taken using the ESO VLT service mode and an observing log is presented in Table~1.

Standard star observations were provided to us, of both polarized standards and unpolarized standards, using the same observing procedures as the target clusters.

\subsection{Data Processing}

A master bias frame was prepared from the median of 55 bias calibration frames. Prescan and overscan rows were used to match the bias level to each data
frame and hence to subtract the master bias. A pixel sensivitity flat field (P-flat) was derived separately for the o- and e-beams using white light dome flats, and dividing each flat field section by its row average, where, to a good approximation, the rows correspond to the wavelength direction. The data were all divided by the P-flat after debiassing.

The 2D subsections of the data arrays corresponding to the two different polarization beams were extracted, and arc line lamp spectra were used to rectify and wavelength calibrate the data. A shear was applied to the data frames to straighten arc lines in the y-direction only, and a second shear was applied to straighten the spectra in the x-direction. Wavelength calibration was assumed to be the same for all observations. We did not resample the data in the wavelength direction, but provided an external lookup table of wavelengths for each pixel. The spatial separation of the two polarization beams was measured using standard star observations. Hence with spline interpolation, from each original data frame, we derived two frames, one for each polarization beam, spatially and spectrally registered. The accuracy acheived is much better than a single pixel, so we anticipate no significant error term from these procedures as line polarization measurements integrate over multiple pixels.

A simple cosmic ray rejection algorithm was applied to the data frames by comparing each spectrum of a target to the median of all similar spectra, typically 16. Two dimensional spectra for each retarder configuration were co-added, separately for the o- and e-beams.

To derive the polarization information, there are two methods available. One is the difference method, and the other the flux ratio method (Miller et al. 1987). We processed the data using both techniques but found no significant difference in the results.
For the flux ratio (FR) method, the normalized Stokes parameters are given by
$q = (R_q - 1)/(R_q + 1)$ where $R_q= \sqrt{(I_0^o/I_0^e)/(I_{45}^o/I_{45}^e)}$
and
$u = (R_u - 1)/(R_u + 1)$ where $R_u= \sqrt{(I_{22.5}^o/I_{22.5}^e)/(I_{67.5}^o/I_{67.5}^e)}$.
For the $O-E$ difference method (OE), the normalized Stokes parameters are given by
$q=0.5((I_0^o- I_0^e))/(I_0^o+I_0^e))-0.5((I_{45}^o-I_{45}^e)/(I_{45}^o+I_{45}^e))$
and
$u=0.5((I_{22.5}^o - I_{22.5}^e)/(I_{22.5}^o + I_{22.5}^e)) - 0.5((I_{67.5}^o-I_{67.5}^e)/(I_{67.5}^o + I_{67.5}^e))$.

The retarder offset angles, as provided at the ESO web site, were also subtracted from the derived position angle data\footnote{\url{http://www.eso.org/sci/facilities/paranal/instruments/fors/inst/pola.html}}.

Adjustable smoothing parameters in both the y (spatial) and x (wavelength)  directions were allowed in the processing but in the end, we
used only data at the highest resolution in order to minimize correlated noise terms.
Where appropriate, a throughput correction was applied, derived from the  FORS total efficiency as provided by the FORS2 exposure time calculator  on the ESO website, interpolated and corrected to account for the slowly changing pixel size with wavelength.

We processed the standard star observations in the same fashion as the data, Table~2, and found excellent agreement with the expected values for the polarization and position angle. Fig.~5 shows the derived polarization and position angle as a function of wavelength for the polarized standard star Vela~1. The agreement with expectation is of order 0.1\%\ in polarization degree, and within $1^{\circ}$ in position angle over most of the spectrum. There is no significant difference between the two reduction methods (OE or FR) for the standards or any of the targets, hence throughout, we describe only the FR method for convenience.
The subsequent analysis steps are illustrated and described within the results sections 3.1 and 3.2.

\section{RESULTS}

\subsection{Synchrotron Emission in M87}

The central region of M87 contains highly polarized synchrotron emission and emission from the nucleus which is likely to be synchrotron in origin (Perlman et al. 2011). This serves as an additional check on our methods while providing  astrophysically interesting results.
The slit positions for M87 are shown in Fig.~1. The slit  passes through the nucleus then extends along the bright emission filament north of the famous synchrotron jet, shown in the upper panel. The lower contrast image in the lower panel shows that the compact source within the jet, HST-1, which underwent a massive ouburst peaking around 2005 lasting several years
(Perlman et al. 2003; Madrid 2009),
is also included in our slit. The separation between the nucleus and HST-1 is only 0.8~arcsec, yet they are clearly and cleanly separated in the spectra,  Fig.~6 inset, illustrating the excellent seeing conditions for these observations.

Fig.~6 shows the overall data processing approach for the example of M87. The basic CCD reductions, geometric corrections and wavelength calibrations lead to separate o- and e-beam images, which are combined using the FR method to yield Stokes q and Stokes u images. To derive a Stokes I image, it was necessary to coadd the polarization spectra and remove night sky emission lines. We did this by first deriving a spatially averaged continuum profile  (y direction) of the galaxy in a region of the spectrum unaffected by emission lines from the sky or from the activity in the galaxy. We also established a mask indicating the location of emission lines within the objects. At each wavelength, the continuum spatial profile was  scaled linearly to the data at that wavelength omitting regions with internal emission lines. The intercept of the linear fit, corresponding to a constant offset, yielded a model of the night sky line distribution which was subtracted from the total intensity image. The resulting image serves as the ``Stokes I'' total intensity image, $I(\lambda, y)$. The scaled model continuum and night sky emission line maps were both subtracted to yield images of the galaxy line emission total intensity, Fig.~6.

Given that the spectra all contain a mixture of sources, including continuum stellar emission from the galaxy, line emission from the galaxy associated with its radio source, and in M87, optical synchrotron continuum emission, it is appropriate to work with the ``polarized flux'' rather than the polarization degree. This allows us to separate the different constituents of the polarization and study the individual polarization of discrete sources separately. We derive the polarized flux, $p_f$, and corresponding total intensity Stokes parameters $Q, U$ as $Q = I q$, $U = I u$ and $p_f = \sqrt{Q^2+U^2} = I p_d$ if $p_d$ is the polarization degree.

To examine the M87 synchrotron sources, we extracted spectra three pixels wide spatially, centered on both the nucleus, and HST-1. With this width, there is no overlap between the two extractions (HST-1 and the nucleus are measured to be 3.13 pixels apart using quadratic fits to spectrally averaged spatial profiles). We will use the complete polarized flux spectra below, when looking at line emission, but to check the nuclear polarization and HST-1 polarization, we used these extracted spectra of the total intensity and polarized flux. We also extracted a galaxy continuum spectrum from a region away from line emission, 3.5 to 7~arcsec SE of the nucleus and linearly scaled and subtracted this from the total intensity spectra. Comparing the mean values of the polarized flux, and the individual Stokes Q, U spectra in the region 550---600~nm to the galaxy-subtracted total intensity levels, we derived a nuclear polarization of 
12.0\%\ and a polarization for HST-1 of 23.4\% . The formal statistical errors are negligible, and we are likely dominated here by systematic errors from the galaxy subtraction process.
The position angle of the nuclear polarization electric vector around $\approx 600$~nm is $128^{\circ}\pm 0.3^{\circ}$ (uncertainty from the measured dispersion of the position angle) while the inner jet has
position angle $\approx 290^{\circ}$ (Cheung et al. 2007),  i.e. a misalignment of $\sim 18^{\circ}$. If the polarization is synchrotron, it is normal to consider the magnetic field
position angle, which is at 90$^{\circ}$ to the electric vector, hence at p.a. $38^{\circ}$, or $\sim 18^{\circ}$ from perpendicular to the jet. HST-1 is undoubtedly synchrotron, and its magnetic vector position angle is $24^{\circ}\pm 0.4^{\circ}$, which is $\approx 94^{\circ}$ from the jet axis,  close to perpendicular to the jet.
For HST-1 Perlman et al. (2011) found a relatively stable magnetic vector position angle of $\sim 28^{\circ}$for HST-1, with polarization degree ranging 20---40\% , and a nuclear position angle (electric vector)  varying wildly between 100$^{\circ}$  and 180$^{\circ}$ and polarization 1--14\%\ hence we are comfortably within this range and in good agreement for HST-1.

\subsection{Line Emission Polarization}

Figs.~7 to 10 show images of line emission spectra and polarized flux spectra. As for the case of the M87 nucleus described above, we scaled and subtracted galaxy continua spectra to derive the line emission spectral images shown.
Also as described above, we use polarized flux rather than polarization degree to separate different components more easily than
polarization degree would allow. For example, again in the case of M87, the polarization degree spectrum dips dramatically at $H\alpha$, but the polarized flux spectrum shows a smooth uninterrupted continuum through this region, Figs.~11---13 discussed below.

To process line emission polarization measurements, we determined a spatial extraction region (range of y values), and then extracted spectra for Stokes I, Q, U and the line emission image (which is the galaxy continuum subtracted Stokes I image). Focussing on a line or group of lines, we then selected the region of the spectrum where the lines are present, and a region either side, and fitted a straight line continuum across the line emission for both the Q and U spectra. Stokes parameters are linear, and hence if there is emission line polarization, this would be revealed as an additive component to the Stokes parameter spectra in the vicinity of the emission line. Hence, the correct procedure to determine whether line emission polarization is present, is to subtract the
underlying continuum of the Stokes Q and U spectra, which was done. 
The estimate of the emission line polarized flux and degree are then:
$$p_f = \sqrt{Q_s^2 + U_s^2} - \Delta p$$
$$p_d = p_f/I_e$$
where $Q_s$ and $U_s$ are the total values of the continuum subtracted Stokes parameters in the emission line region,  $I_e$ is the total value of
Stokes I in the emission line region, and $\Delta p$ is a bias term, described below. As before, the position angle is ${1\over 2} tan^{-1}(U_s/Q_s) + \phi$ where $\phi$ is the spectrograph slit
position angle on the sky.

For low values of polarization, the positive definite nature of the polarized flux causes a bias towards positive values. If the true polarization is zero, then the expected value of the polarized flux is $ \Delta p_c = \sqrt{\sigma_Q^2+\sigma_U^2}$ where $\sigma_Q$ and $\sigma_U$ are the uncertainties on $Q$ and $U$. We estimated $\sigma_Q$ and $\sigma_U$ empirically from the root mean square of the residuals from the straight-line continuum fits. Within the line emission region, the count level can be extremely high, and hence we adjusted the derived uncertainties assuming Poisson counting statistics, scaled from the empirical continuum level. That is, the assumed bias on the polarized flux is $\Delta p =  \sqrt{\sigma_Q^2+\sigma_U^2}\sqrt{I_e/I_c}$ where $I_c$ is the continuum Stokes I and $I_e$ is the total flux in the emission line.
Figs.~11---13  illustrate for the example of the M87 nucleus $H\alpha+[NII]$ complex. Fig~11 shows the total flux per pixel for the M87 nucleus
in the vicinity of the $H\alpha+[NII]$ lines, Stokes I. The solid vertical line shows the center of the complex, and the dotted lines, the bounds used to define
the location of the line emission. The (red) crosses indicate the continuum region that was used for fitting purposes.
Fig.~12 shows the Stokes parameters Q and U. An arbitrarily scaled Stokes I is included for reference and also the scaled polarization degree showing the strong dip at the locations of the emission lines. For the nucleus of M87 the continuum is highly polarized, and well-described as a power-law continuum, see above.
We fitted straight lines using the regions indicated by red crosses to the Stokes Q and U and subtracted the continua as described above.
The results of this subtraction are shown in Fig.~13, which also illustrates the necessity to remove the bias of the polarized flux due to its positive definite character. The upper red line shows the unbiased polarized flux, i.e. the quadratic sum of the continuum-subtracted Q and U spectra. If the line polarization is zero, then we expect, using the procedure described above to derive a noise model, the dotted red line shown. Clearly
this mimics the behaviour of the data quite closely.
To correct the polarized flux spectrum, we therefore subtract this bias term from the polarized flux, producing the solid black line, which effectively removes the apparent polarized flux excess at the location of $H\alpha+[NII]$.
The final derived polarization is $p=0.00038 \pm 0.00033$ for this line complex.

This is the generic procedure followed to populate the primary results presented in Table~3.
Table~3 includes polarization measurements for all slit positions in all target objects for the $H\alpha+[NII]$ complex, a narrow $H\alpha$-only region selected to be 2~nm width centered on $H\alpha$, and the strongest lines which are [OIII]~5007, [OI]~6300, and [SII]~6717+6731.Table~3 presents a summary of the results, with polarization upper limits for all strong lines in all objects. 
Since all measurements are upper limits, we do not present derived position angles, since they would be meaningless.

\subsubsection{M87}

Fig.~6, inset, shows  dramatically the two synchrotron components discussed in \S~3,1 and visible intersecting the slit in Fig.~1. The two highly polarized components are cleanly separated, with a separation of 0.8~arcsec. 
Fig.~7 shows the line emission spectrum for the M87 slit 1 position, with the corresponding polarized flux spectrum at two different contrast levels.
There are no obvious features at the location of the emission lines, indicating that our procedure for deriving the polarized flux is sound.
Fig.~7 also shows the polarized flux at extremely high contrast, and there is a hint of polarized emission revealed at the location of the strongest lines.
However, given the positive definite character of the polarized flux, as described above, this is due to enhanced
noise from counting statistics in a region of high intensity.
The spectra shown in Fig.~7 were extracted from a 3~pixel wide region centered on the nucleus, and as already shown, result in  an upper limit to the polarization, while the polarized flux spectrum  is smooth and continuous through this region.

To measure the polarization in the extended line emission region visible in Fig.~1 and Fig.~7, upper panel, for the slit passing through the nucleus, we extracted a spatial region extending from just beyond HST-1 to the edge of the visible emission. This filament is approximately radial, diverging from the jet with distance from the nucleus. It is slightly blueshifted, and the presence of dust absorption in the filament core shows it to
be in the front side of the galaxy, hence the material may be in outflow (Sparks, Ford and Kinney 1993).
By eye, from Fig.~7, there is no apparent polarized light, and the values provided in Table~3 quantify this. 

Slit 2 passes through the position of the HST/COS FUV observation described in Sparks et al. (2012). The FUV spectrum shows CIV 1550 and HeII 1640 line emission at a level consistent with their arising from a conduction interface between the cool filament and surrounding reservoir of hot gas.
The fiducial conductive models used in Sparks et al. (2009); Sparks et al. (2012) were found not to be in the saturated regime, to which we return in the following section. The corresponding optical polarization spectra are shown in Fig.~7, middle panels.

Slit 3 passes through the complex of $H\alpha$ emission that is present where infalling line-emission filaments encounter the SE radio lobe `behind' the plane of the sky through the nucleus (Sparks, Ford \&\ Kinney 1993). This emission is Arp's (1967) counterjet. The velocity field of the filaments beyond this location is blueshifted, and the SE radio lobe is thought to be the more distant of the two inner lobes. Hence, if the filament is physically associated with the SE lobe, as it appears to be, it is most likely infalling. The polarization spectra are shown in Fig.~7, and the derived upper limits listed in Table~3.

\subsubsection{Hydra~A}

The acquisition image of Hydra~A, Fig.~4, shows a compact, edge-on, prominent dust lane through the center of the galaxy, running in position angle $\approx 100^{\circ}$. A lobe of emission at the west end of the dust lane further distorts the galaxy contours. The dust lane also marks the location of a rotating line-emitting disk (Heckman et al. 1989). The long slit runs in position angle 20$^{\circ}$, approximately along the axis of the large-scale radio jet. The long slit line emission spectrum, Fig.~8, shows strong lines at the location of the nucleus and dust lane,with a patch of emission to the north. The presence of dust distorts the background appearance of the line emission 2D spectrum, though this does not affect our conclusions.

The polarized flux data, Fig.8, show elevated levels of apparent polarization at the position of the dust lane, and also excess emission at the positions of $H\alpha+[NII]$ and $[SII] 6717, 6731$. However, as shown in Fig.~14, this also is a consequence of the positive definite nature of the
polarized flux in the presence of higher noise due to Poisson counting statistics. The derived position angle is quite close to that of the dust lane, but again the uncertainties are too large to allow us to conclude we have a definitive measurement of polarization, formally only $~1.25\sigma$ for $H\alpha$ on the nucleus. We processed the nucleus and the patch of emission offset from the nucleus separately, and list both in Table~3.

\subsubsection{NGC~4696}

The long slit emission line spectra for NGC~4696 are shown in Figs.~2 and Fig.~9. The two long-slits are parallel to one another. One passes through the nucleus, and the other through the prominent, extended dust lane which is mostly coincident with the $H\alpha$ emission-line filaments.
The polarization spectra are shown for both positions in Fig.~9. 
As expected we see strong, low-excitation line emission from the nucleus, extended to the NW. We also see the expected strong, low-excitation line emission from the region of the dust lane. The nucleus spectrum also shows a deficit of emission coincident with NaD absorption, as discussed by Sparks et al. (1997). There is no significant polarization associated with either of these two spectra, Table~3.

\subsubsection{PKS0745-19}

The strong emission lines of PKS0745-19 are readily apparent in the spectra displayed as Fig.~10.  The overall appearance of the spectra is very similar to that of Hydra~A, even to the presence of a compact nuclear absorbing dust lane. As in the case of Hydra~A, the apparent presence of slightly polarized emission is attributable to the random errors, and no significant line polarization was derived for this object, in either of the two slit positions, Table~3.

Table~3 presents a summary of the results, with polarization upper limits for all strong lines in all objects. 
 
\section{DISCUSSION}

\subsection{Heating of filaments}

Many different sources of energy and excitation  mechanisms have been considered to drive the $H\alpha$ emission filaments in cool core galaxy clusters. For NGC~1275 (Sabra et al. 2000; Conselice, Gallagher \& Wyse 2001) and M87  (Sabra et al. 2003) photoionization by the central AGN, the intracluster medium,  young hot stars and shock heating have all been discussed as the underlying physical mechanisms involved. The conclusion was that ``neither shocks nor photoionization alone can reproduce the emission line intensity ratios'' and that some additional source of heating must be present. A study of the optical line ratios in Abell 2597 led Voit \& Donahue (1997) to rule out shocks as an excitation mechanism, and to conclude that although hot stars might be the best candidate for producing the ionization, even the hottest stars could not power a nebula as hot as observed, and that another non-ionizing source of heat must contribute at least a comparable amount of power. Similar conclusions were reached by Hatch, Crawford \& Fabian (2007) for a number of cool core clusters, and they also note that heating by thermal electrons from the intracluster medium is a plausible mechanism. 

\subsubsection{Conduction}

Despite early claims to the contrary (e.g. B\"ohringer \& Fabian 1989), it has long been recognized (Sparks et al. 1989) that an important source of heating is likely to be heat flow (thermal conduction) from the  hot $\sim 10^7$K X-ray-emitting gas which makes up much of the intra-cluster medium surrounding the filaments. The plausibility that thermal conduction can play a major role in heating the filaments has been underlined by the finding by Sparks et al. (2009; 2012) that there is $\sim 10^5$ K gas spatially associated with the H$\alpha$ filaments in M87. They find that the measured emission-line fluxes from triply ionized carbon (CIV 1549 \AA) and singly ionized helium (HeII 1640 \AA) are consistent with a simple model in which thermal conduction, using Spitzer conductivity, determines the interaction between the hot and cold phases (Sparks et al. 2009; 2012). 

\subsubsection{Saturated conduction}

It has further been noted that in the tenuous intragalactic medium, where there is a large temperature difference between the medium $\sim 10^7$ K and the filaments $\le 10^4 K$,  the electron mean free paths might be sufficiently large that standard diffusive (Spitzer) conductivity is no longer applicable. Under these circumstances the conduction becomes``'saturated'' at a value around the maximum heat flux in a plasma of order (Cowie \& McKee 1977)
\begin{equation}
\label{sat0}
q_{\textrm sat} \approx  f \frac{3}{2} (n_e k T_e) v_{\textrm char},
\end{equation}
where $n_e$ and $T_e$ are the electron number density and temperature, respectively, and $v_{\textrm char}$ is a characteristic velocity which one might expect to be of order the electron thermal velocity $v_e = \sqrt{3kT_e/m_e}$, where $m_e$ is the electron mass. This is because when conduction reaches its saturated limit, the electrons no longer diffuse (short mean free path) but rather are able to stream freely (mean free path larger than the characteristic local temperature distance scale, $T/ \mid \nabla T \mid $). The reduction factor $f\approx 0.4$  (Cowie \& McKee 1977) accounts for a charge neutrality requirement, which we discuss below.
Sparks et al. (2004) have noted that saturated conduction, using the formula (Cowie \& McKee 1977)
\begin{equation}
\label{sat1}
q_{\textrm sat} = 0.4 \left( \frac{2kT_e}{\pi m_e} \right)^{1/2} \: n_ekT_e,
\end{equation}
can provide an adequate heat flux to power the filaments in both M87 and also NGC 1275. Fabian et al. (2011) concur that the surface radiative flux from the outer filaments in NGC 1275 is close to the energy flux impacting on them from particles in the hot gas. They use a different formula for the saturated conduction heat flux, also given by Cowie \& McKee (1977),
\begin{equation}
\label{sat2}
q_{\textrm sat} = 5 \, \phi \, p \, c_s,
\end{equation}
where $\phi \sim 1$ accounts for uncertain physics such as $f$ and the average particle mass, $p = nkT$ is the gas pressure (where here $n$ is the total particle density and $T$ the temperature) and $c_s$ is the isothermal sound speed of the gas, $c_s = \sqrt{p/\rho} \sim \sqrt{kT/m_p}$, where $m_p$ is the proton mass. 

It is worth remarking that these these two formulae (equations~\ref{sat1} and~\ref{sat2}) emphasize physically different ways of viewing the conduction process, though they are algebraically equivalent. In equation~\ref{sat1} the characteristic velocity (equation~\ref{sat0}) is taken to be the free electron speed averaged over direction, of order the electron thermal velocity
\begin{equation}
v_{\textrm char} = \left( \frac{8}{9\pi} \right)^{1/2} \left( \frac{kT}{m_e} \right)^{1/2}.
\end{equation}
In equation~\ref{sat2} the characteristic velocity which is implied is the ion (or proton) sound speed $v_{\textrm char} \sim v_i \sim (m_e/m_p)^{1/2} v_e$.

In fact the equations are equivalent for the case of equal ion and electron temperature, and the actual characteristic velocity with which electrons and ions cross the boundary is the same. To see this, imagine that initially one sets a hot fully ionized plasma (with large mean free path) next to a cold absorber. Then initally, since $v_e \gg v_i$ there is a flow of heat from the hot plasma at a rate given by equation~\ref{sat0}.
However, this flux of heat, carried by the electrons, leads to a net electric current $j$ into the absorber, and therefore to a build up of (negative) charge on the absorber. What then happens is that ``an electrostatic field $E$ will build up to such a value that $j$ vanishes. This field then reduces the flow of heat'' (Spitzer \& H\"arm, 1953). Eventually, in order to maintain the zero current condition, the net flow speed of both the electrons and the ions when they reach the cold absorber must be the same. Thus the electrostatic field set up on the absorber is such that it slows the electrons and speeds the ions. An electron loses the same amount of energy when it travels through an electrostatic potential barrier as a proton gains when it falls into an electrostatic potential well. Because $m_p \gg m_e$, the net flow speed of both the protons and the electrons must be of order $v_{\textrm char} \sim v_i \sim c_s$, in line with the expression for saturated heat flux given by equation~\ref{sat2}.

This now has a further very important implication. Because, to maintain charge neutrality, the net flow speeds of the electrons and the ions must be the same, the dominant {\it energy transport is provided by the ions}. 

\subsection{Theoretical Degree of Polarization}

We have seen that if the dominant heating mechanism for the filaments is indeed the penetration of the filaments by thermal particles originating in the hot gas (Fabian et al. 2011), then most of the energy is carried by the hot ions. One effect of the excitation of H atoms (and H molecules) by a non-isotropic velocity distribution of electron or protons is that the resultant emission lines (including Ly$\alpha$ and H$\alpha$) can be polarized. This has been discussed in the context of excitation by electron impact in solar flares (Laming 1990a; Aboudarham et al. 1992) and for excitation by both electron and proton impact in non-radiative shock fronts to be found in supernova remnants (Laming 1990b). To understand how the polarization comes about, consider a collection of H atoms  excited by a beam of protons. There is  a preferred plane perpendicular to the velocity vector of the beam. As viewed from that plane, and for the relevant range of proton energies in the hot gas surrounding the filaments (a few keV), it is not trivial to calculate the expected emission line polarization. 

At such energies the contribution to line polarization for electron impact excitation in H$\alpha$ is negligible (Aboudarham et al. 1992). Therefore, if heating of the filaments were to proceed via standard diffusive (Spitzer) conductivity, in which most of the heat is carried inwards by electrons,  we would expect negligible polarization of the emission lines.
For proton energies of a few keV, polarization can arise, however it is not straightforward to calculate the  precise level.
The difficulty arises mainly because at such energies the proton velocity,
\begin{equation}
v_p = 4.34 \times 10^7 (E/keV)^{1/2} cm/s
\end{equation}
is much less than the electron orbital velocity in the ground state
\begin{equation}
v_e = \frac{ \hbar c}{4 \pi^2 e^2} \, c  = 2.2 \times 10^8 cm/s,
\end{equation}
so that the plasma is in the quasi-molecular regime where electronic processes proceed through states which are transiently formed during the collision (Hippler et al. 1988)\footnote{Note that these protons are generally not able to ionize the H atom. This is because, since $m_p/m_e \gg 1$, the change in electron velocity caused by a collision between an electron and a proton is of order $v_p$. For a ground state electron, $v_e \gg v_p$ and so it gains little energy from the proton. In contrast, an incoming electron with velocity $v \sim v_e$ changes the velocity of an electron it collides with by of order $v$. Thus energy exchange with another electron is very efficient, making it easier to ionize an atom with an electron than with a proton of the same energy; see, for example, Lin et al. (2011).}.

Both computational and experimental results for Ly$\alpha$ suggest polarizations in the range of 10\%\ --- 25\%\  (Kauppila et al. 1970; Hippler et al. 1988; McLaughlin, Winter \& McCann, 1997; Keim et al. 2005). Laming (1990b) suggests that at these energies it is appropriate to assume that the H$\alpha$ polarization is the same as the Ly$\alpha$ polarization for the same energy photons. Computations by Balan\c{c}a \& Feautrier (1998) indicate that this is an appropriate assumption and also find that proton impact polarizations for H$\alpha$ for protons in the range $E \approx$ 1 -- 5 keV are of order 20\%\ --- 25\% .

\subsection{Realistic predictions for polarization}

If the $H\alpha$ emission lines from the filaments are  being induced by hot particles originating in the hot gas, then the observed polarizations are likely to be less than these values. The reduction would come about because the incoming particles do not form an organized beam, because of geometric effects, and because the cross-section for excitation to the $n=3$ level (in order to excite H$\alpha$) can be comparable in this energy range to the cross-section for ionization (Lin et al. 2011). 
We briefly consider geometry, and two mechanisms which have the potential to randomize the proton velocity distribution within the filaments. We conclude that measurable polarization ought to persist for the saturated conduction case.

\subsubsection{Geometry}

Because of the approximately cylindrical nature of the filament geometry, and because the electrostatic field enhances the anisotropy of the proton velocity distribution close to the interface by accelerating protons towards it, it is difficult to envisage geometric factors reducing the polarization by even as much as an order of magnitude. For geometry to negate polarization, a highly  contrived geometrical configuration would be required. For example, if all the gas in the filaments were to be in the form of  spherical clouds, the symmetry with respect to angle on the sky would cancel the polarization. The linear,  filamentary morphology, however, suggests that such a topology is unlikely. 
If the emission line filaments consist of many strands, or ``threads'' (Fabian et al. 2008), then the larger filament would exhibit Stokes parameters which are the average of the individual strands. If these were completely disordered, the polarization could be reduced or eliminated, however since individual strands are observed to align in order to produce the macroscopic filament structure, the polarization would be similar to the polarization of a single cylinder and would not cancel.
There may be regions, such as those close to the nuclei of the galaxies, where we do have a mixture of filament directions along the line of sight, and the consequent averaging could contribute to a dilution of the average polarization. Most of the filament regions are, however, relatively well-ordered and we would anticipate that the polarization would largely be preserved.
Thus, if saturated conduction is the dominant excitation mechanism for the emission lines from the filaments, we may expect the lines to be polarized at least at the level of a few per cent, even taking geometric effects into account.

\subsubsection{Scattering}

If the proton beam is scattered so that the proton velocities get randomized, then this could significantly reduce the degree of polarization. To excite the electron from the $n=1$ state to $n=2$ or $n=3$, the proton needs to come within a few Bohr radii $a_0$, where
\begin{equation}
a_0 = \frac{h^2}{4 \pi^2 m_e e^2} = 0.53 \times 10^{-8} cm.
\end{equation}
This value agrees with the typical cross-sections for the interaction given as  of order $\sim 2 \times 10^{-17}$ cm$^2$ (e.g. McLaughlin et al. 1997; Balan\c{c}a \& Feautrier, 1998; Lin et al. 2011), compared to the area of the first Bohr orbit of $\pi a_0^2 = 8.8 \times 10^{-17}$ cm$^2$. At the radius of the Bohr orbit, the electrostatic potential energy is, of course, around $E_0 \approx 13.6$ eV which is much less than the typical proton energies $E$ we are interested in, of a few keV. Thus the angle through which the proton is deflected is of order $\phi \sim E_0/E \ll 1$.  We conclude that the act of exciting the Lyman and Balmer lines does not significantly isotropize the directions of the incoming ions.

\subsubsection{Magnetic fields}

For typical magnetic field strengths expected within the filaments, the Larmor radius for a few keV proton is of order $10^9$ cm, which is many orders of magnitudes less than the radii of the filaments. Hence the proton velocity distribution could potentially be isotropized if the magnetic field structure within the filaments were strongly randomized.

Fabian et al. (2008) have argued, however, that the filaments in NGC~1275 are ``essentially magnetic structures'' in which the magnetic pressure dominates the thermal pressure. Werner et al. (2013) came to similar conclusions for the filaments in M~87. The suggested value of $B \approx 100 \mu$G would give approximate pressure equilibrium with the external medium (density $n \approx 0.06$ cm$^{-3}$ and temperature $T \approx 4$ keV) and would imply for their assumed values internal to the filament of density $n \approx 2$ cm$^{-3}$ and temperature $T \le 10^4$ K that the ratio of thermal to magnetic pressure $\beta$ is:
\begin{equation}
\beta = \frac{nkT}{B^2/8\pi} \le 7 \times 10^{-3}.
\end{equation}

	Similar arguments are made for the filamentary gas in M~87 by Werner et al. (2013) who suggest that the $10^4$ K gas phase, which emits the density sensitive [SII] $\lambda$ 6716, $ \lambda$6731, requires fields $B \approx 50 \mu$G to maintain pressure balance with the surroundings.

In this picture, it is argued that the magnetic field must lie predominantly along the filaments, in order that they be magnetically dominated structures. Note that Fabian et al. (2008) also deemed it necessary that there is an ``unseen''~\footnote{It is not clear what ``unseen'' means in this context, since none of the magnetic fields mentioned in Fabian et al. (2008) are actually observed -- they are simply  hypothesized to exist in order to maintain the assumed structure of the filaments.
} component of magnetic field which is perpendicular to the filaments in order to prevent material sliding along the filaments. It is not clear how both these two requirements are to be achieved simultaneously. In addition it is claimed that the filaments contain Alfv\'{e}nic turbulence in order to account for the internal velocity dispersion of $\sim 100$ km s$^{-1}$ (Hatch et al. 2006). The driver for this turbulence remains unspecified.

	The idea of turbulence within the filaments is also suggested by Fabian et al. (2011) and Werner et al. (2013). There, in order that external hot plasma can interpenetrate the cold filaments, a process known as ``reconnection-diffusion'' is introduced. 
But these authors agree that (Fabian et al. 2008)  ``it is natural to assume that the turbulent velocity in filaments is less than the Alfv\'{e}n speed'', because otherwise the turbulence would randomize the field direction and so prevent the existence of long-lived filaments. This is the crux of the matter for our discussion. In order for the filaments to be strongly magnetic it is necessary that the magnetic field within the filaments be well-ordered. If the field is well-ordered, magnetic randomization of the velocity distribution of the incoming protons is not going to be effective. Thus the reduction of polarization caused by randomization of the proton velocity distribution by a chaotic magnetic field configuration is unlikely to be significant.

\subsection{Implication of the Observations}

Our observational limits are very stringent. For individual emission regions, we find polarization levels $\lesssim 0.1\%$, with the average polarization degree  in Table~3 for the $H\alpha+[NII]$ complex $<p>\approx 3\times 10^{-4}$, i.e. a polarization percentage $\approx 0.03\%$.
By contrast, from theoretical considerations, we have argued that in a saturated conduction regime where the filaments are excited by a highly directional proton beam, polarization levels of plausibly a few percent ought to be present in the emission lines.
This arises because the protons do not necessarily ionize the filament neutral H atoms, and the system retains a degree of the incident anisotropy.

Hence, we conclude that the evidence from these observations and theoretical arguments, is that if conduction is the dominant process
for energy transport into the filament system from the hot ambient coronal X--ray gas, we are unlikely to be in the saturated regime.
For the case of M87, Sparks et al. (2012) showed that the line strengths were consistent with a classical non-saturated conduction
model, which would not be expected to produce significant polarization.
Global energetic considerations do show that saturated conduction can carry the required energy to power the emission filaments, and
to order of magnitude the energy transport is similar for the classical conduction regime, though the details of the interface structure, energy flux and timescales involved  differ.

The use of emission line polarization as a plasma diagnostic is clearly in its infancy for application to galaxy clusters.
From other areas of astrophysics, it is apparent that the approach has the potential to provide unique insights into the excitation
mechanisms of relevance.
Additional theoretical work is needed to determine more accurately the likely levels of polarization, not just for the case of conduction, 
but also for anisotropic photon excitation and shocks.
Heuristically, one would expect that for photoionization, the resulting polarization distribution will depend primarily on the degree of anisotropy of the photons, both their origin and modification by any dust present, as well as, of course, on the spectral energy distribution of the photons.
Plausibly, hot stars and the intracluster medium would result in an approximately isotropic photon excitation and low polarization, while AGN excitation would have much stronger directionality and is therefore more likely to yield polarization.
Shock excitation is also highly directional and the consequent polarization  depends on the shock speed. Laming (1990b) showed
that substantial polarization can arise from fast shocks, of order 2000~km/s, which is much faster than likely shock speeds in these
filaments, unless an interaction with, e.g., the relativistic plasma radio lobes is involved.

\section{Conclusion}

Motivated by the potential of an innovative new diagnostic applied to galaxy cluster physics, we have acquired deep long slit spectropolarimetry of the low excitation filament systems in four cool core clusters. Polarimetry of line emission can in principle distinguish between several competing
forms of excitation and hence help ascertain the transport processes that govern the physical characteristics and evolution of gas in galaxy clusters.
We detected the expected levels of polarization for two synchrotron sources  in M87, the nucleus and jet knot HST-1, validating our observational aproach and offering a useful check on previous imaging polarimetry of these sources.
Two of the galaxies have edge-on nuclear dust lanes, and superficially show a slight excess of polarization. Nuclear polarization in such cases may plausibly be attributable to  dichroic absorption through aligned grains or scattering into the line
of sight of a hidden AGN. Formally, however, the magnitude of this polarization is not significant.
All emission lines, both on the nucleus and in the extended low excitation emission regions, show polarization upper limits at levels of order 0.1---0.05\%.

There seems to be a growing consensus that the heating of the low-excitation $H\alpha$ filaments found in cool core clusters is achieved by some form of thermal conduction. Sparks et al. (2012)  successfully modelled the excitation of CIV $\lambda 1550$ and HeII $\lambda$ 1640 in the filaments in M~87 using standard (unsaturated) Spitzer diffusivity in which the energy is carried predominantly by hot electrons. If this is also the source of excitation of $H\alpha$, then the $H\alpha$ lines would, as observed,  be expected to display negligible polarization.

By contrast, Fabian et al. (2011)  argued that the dominant heating mechanism is penetration of the filaments by thermal particles originating in the hot gas, i.e. saturated thermal conduction. In this case, however, because the particle flux is strongly anisotropic, and if the $H\alpha$ lines are excited predominantly by these particles, then the lines are expected to be linearly polarized. For a fully ordered particle beam, at the appropriate energies, the degree of polarization is expected to be high ($\sim 20$ per cent). We have argued,  \S~4.3, that it is unlikely that simple geometric or momentum-redistribution effects would reduce this prediction by much more than an order of magnitude. It is difficult to be more precise than this in the 
absence of a more detailed model for the emission line excitation process, or knowledge of the filament topology and its magnetic field structure.

We have found that the $H\alpha$ lines for all four target clusters have fractional linear polarizations less than an upper limit of around 0.05\%\ --- 0.1\% . The straightforward conclusion to draw from this is that the $H\alpha$ emission lines are not excited by a simple beam of non-thermal particles originating in the hot gas, but variants such as classical Spitzer thermal conductivity are viable.

Though this initial foray into the diagnostic suite afforded by line emission polarimetry produced only upper limits, the long term potential of
the observational approach is substantial. With a mixture of emission line types such as forbidden, permitted, and resonance responding differently to
different modes of excitation, such as collisional ion, collisional electron, photoionization and in different energy ranges and conditions such as saturated or unsaturated conduction, shocks, and  highly directional photoionization, ultimately, the power to distinguish competing
physical transport processes may be unparalleled. Taken in conjunction with spectroscopic models spanning the range
of temperatures known to be present, these diagnostics may eventually reveal the physical processes operating in the  galaxy cluster ISM and their evolution over time and in a variety of situations.




\acknowledgments

Based on observations collected at the European Organisation for Astronomical Research in the Southern Hemisphere, Chile under program 086.B-0138A.
We acknowledge support from grants HST GO-12271 and GO-11681.  STScI/AURA is operated under grant NAS5-26555. JEP thanks the Distinguished Visitor Program at STScI for its continued hospitality.



{\it Facilities:} \facility{ESO (VLT)}.

\bibliographystyle{apj}                       


\clearpage

\setlength{\tabcolsep}{0.04in}
\begin{table}
\begin{center}
\caption{Observing log.\label{tbl-1}}
\begin{tabular}{lccccccc}
\tableline\tableline
Target & Total exposure (s)&ObsID&Date &Slit p.a. & R.A. (J2000)& Dec (J2000) \\
\tableline
PKS0745-19 & 9664 & 496810 & 2010-10-13 & 90.0 & 7:47:31.4&-19:17:40.7 \\
 & & 496812 & 2010-10-12&  & &  \\
PKS0745-19 & 9664 & 496807 & 2010-11-06 & 45.0 &7:47:31.3 &-19:17:41.4 \\
 & & 496809& 2010-11-06& & &  \\
Hydra A & 9856 &  496804 & 2011-01-27 & 20.0  &9:18:05.6  &-12:05:45.0\\
 & & 496806& 2010-12-13& & &\\
NGC4696-dust & 9408 & 496798 & 2011-01-07 & -68.0&12:48:49.2  &-41:18:45.8\\
 & &496800 & 2011-02-10 &  & &&\\
NGC4696-nuc &9408& 496801 &  2011-02-28 &-68.0 &12:48:49.7&-41:18:41.0 \\
 & & 496803& 2011-02-28 & &\\
M87-nuc &9760&  496819 & 2011-02-13& -44.0 &12:30:49.3 &12:30:49.3\\
 & & 496821 & 2011-02-13&  & & &\\
M87-SE &9760& 496813 & 2011-02-08& -38.0 &12:30:51.1 &12:23:25.1\\
 & &496815 & 2011-02-12& &&& \\
M87-fil &9760& 496816& 2011-02-13& -48.0 &12:30:51.3 &12:23:11.4 \\
 && 496818& 2011-02-12&& && \\\tableline

\end{tabular}
\tablecomments{Slit segments are 22 arcsec in length, and slit width 2 arcsec used throughout. Half wave retarder angles used 0, 22.5, 45, 67.5 degrees. Exposure time is divided equally between OB sets for a given pointing.}
\end{center}
\end{table}

\setlength{\tabcolsep}{0.05in}

\begin{table}
\begin{center}
\caption{Polarization standard star observations.\label{tbl-2}}
\begin{tabular}{lcccrrrr}
\tableline\tableline
Target & Wavelength&Polarization \% &p(FR)\% &p(OE)\% &p.a.&p.a.(FR)& p.a.(OE)\\
\tableline
Vela 1&U&6.59&6.51&6.48&169.8&171.3&171.3\\
&B&7.55&7.45&7.45&173.8&170.8&170.8\\
&V&8.24&8.22&8.21&172.1&172.3&172.3\\
&R&7.89&7.97&7.97&172.1&172.1&172.1\\
&I&7.17&7.27&7.25&172.2&172.1&172.1\\
HD 42078&500--700 &0.0&0.26&0.26&\\
HD 97689&500--700 &0.0&0.16&0.16&\\
WD 1620-391&500--700&0.0&0.24&0.24&\\\tableline
\end{tabular}
\end{center}
\end{table}


\begin{deluxetable}{lrrrrrrrrrr}
\tabletypesize{\footnotesize}
\rotate
\centering
\tablecolumns{11}
\tablewidth{0pt}
\tablecaption{Emission line polarization results.\label{tbl-3}}
\tablehead{
\colhead{Target} & \colhead{$p([OIII])$\tablenotemark{*}} & \colhead{$\sigma_p([OIII])$} & \colhead{ $ p([OI])$ } & \colhead{  $\sigma_p([OI])$ } & \colhead{  $ p([H\alpha])$} & \colhead{ $\sigma_p(H\alpha)$} & \colhead{ $ p(H\alpha+[NII])$} & \colhead{ $ \sigma_p(H\alpha+[NII])$} & \colhead{ $ p([SII])$} & \colhead{ $ \sigma_p([SII])$}}
\startdata
M87 nucleus &  0.00056 &  0.00078 &  0.00079 &  0.00052 &  0.00038 &  0.00033 &  0.00023 &  0.00058 & 
 0.00050 &  0.00052\\
M87 extended &  0.00071 &  0.00047 &  0.00002 &  0.00036 & -0.00002 &  0.00024 & -0.00020 &  0.00050 & 
-0.00002 &  0.00022\\
M87 slit 2 &  0.00081 &  0.00085 & -0.00008 &  0.00060 & -0.00030 &  0.00049 & -0.00064 &  0.00099 & 
 0.00065 &  0.00062\\
M87 slit 3 &  0.00012 &  0.00101 & -0.00003 &  0.00094 &  0.00016 &  0.00054 & -0.00062 &  0.00104 & 
 0.00047 &  0.00080\\
Hydra A nucleus & -0.00127 &  0.00160 & -0.00032 &  0.00148 &  0.00110 &  0.00088 &  0.00136 &  0.00150 & 
 0.00236 &  0.00108\\
Hydra A extended & -0.00051 &  0.00358 &  0.00164 &  0.00277 & -0.00095 &  0.00167 & -0.00175 &  0.00338 & 
-0.00074 &  0.00249\\
NGC4696 nucleus &  0.00033 &  0.00134 & -0.00043 &  0.00103 & -0.00012 &  0.00063 & -0.00014 &  0.00124 & 
 0.00080 &  0.00079\\
NGC4696 extended  &  0.00037 &  0.00118 &  0.00021 &  0.00076 &  0.00014 &  0.00048 & -0.00086 &  0.00100 & 
-0.00015 &  0.00063\\
NGC4696 dust  &  0.00032 &  0.00119 &  0.00000 &  0.00079 &  0.00069 &  0.00558 & 0.00091 &  0.00114 & 
0.00015 &  0.00057\\
PKS0745-19 n90 &  0.00445 &  0.00460 & -0.00125 &  0.00238 &  0.00024 &  0.00142 &  0.00068 &  0.00221 & 
 0.00031 &  0.00167\\
PKS0745-19 e90 &  0.00398 &  0.00546 &  0.00096 &  0.00165 &  0.00213 &  0.00084 &  0.00089 &  0.00159 & 
 0.00143 &  0.00115\\
PKS0745-19 n45 &  0.00346 &  0.00340 &  0.00025 &  0.00265 &  0.00082 &  0.00129 &  0.00118 &  0.00198 & 
 0.00228 &  0.00162\\
PKS0745-19 e45 &  0.00144 &  0.00288 &  0.00122 &  0.00193 &  0.00239 &  0.00114 &  0.00316 &  0.00215 & 
 0.00118 &  0.00128\\
\enddata
\tablenotetext{*}{If the debiassing procedure resulted in a negative value of the polarized flux this is retained as a negative value for polarized degree in the table.}
\end{deluxetable}


\clearpage



\clearpage


\clearpage


\begin{figure}
\includegraphics[width=6.5in]{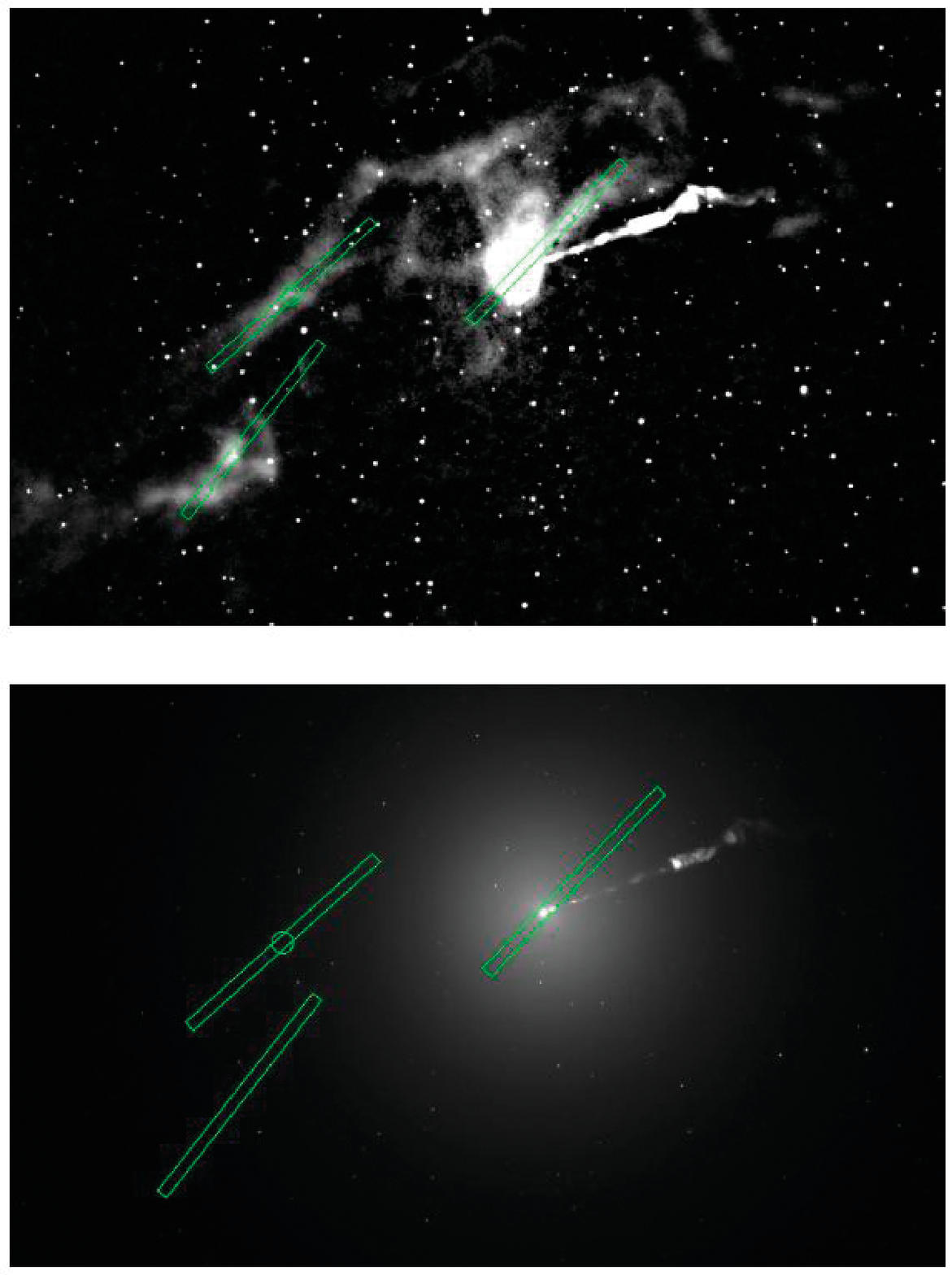}
\caption{Representation of slit positions on M87. The circle indicates the location of the HST/COS aperture used for Sparks et al. (2012).\label{fig1}}
\end{figure}

\begin{figure}
\includegraphics[width=6.5in]{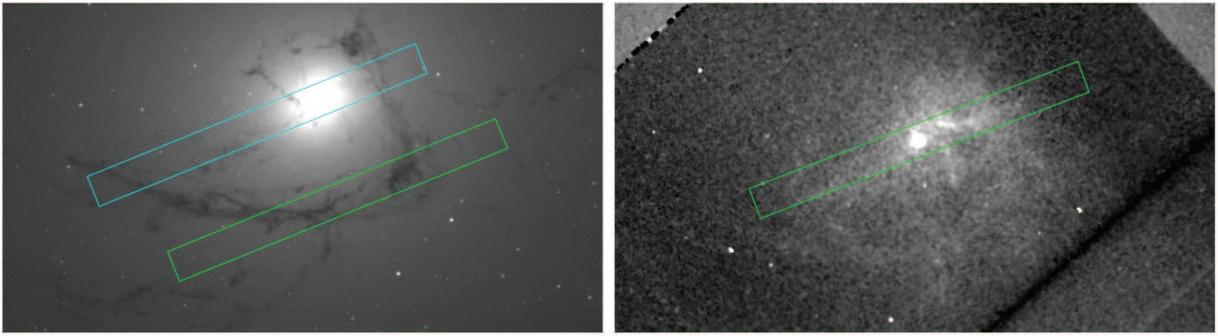}
\caption{Slit positions overlaid on dust (left) and line emission (right) images of NGC~4696. (Only the slit passing across the galaxy nucleus  is shown on the line-emission image.)\label{fig2}}
\end{figure}

\begin{figure}
\includegraphics[width=6.5in]{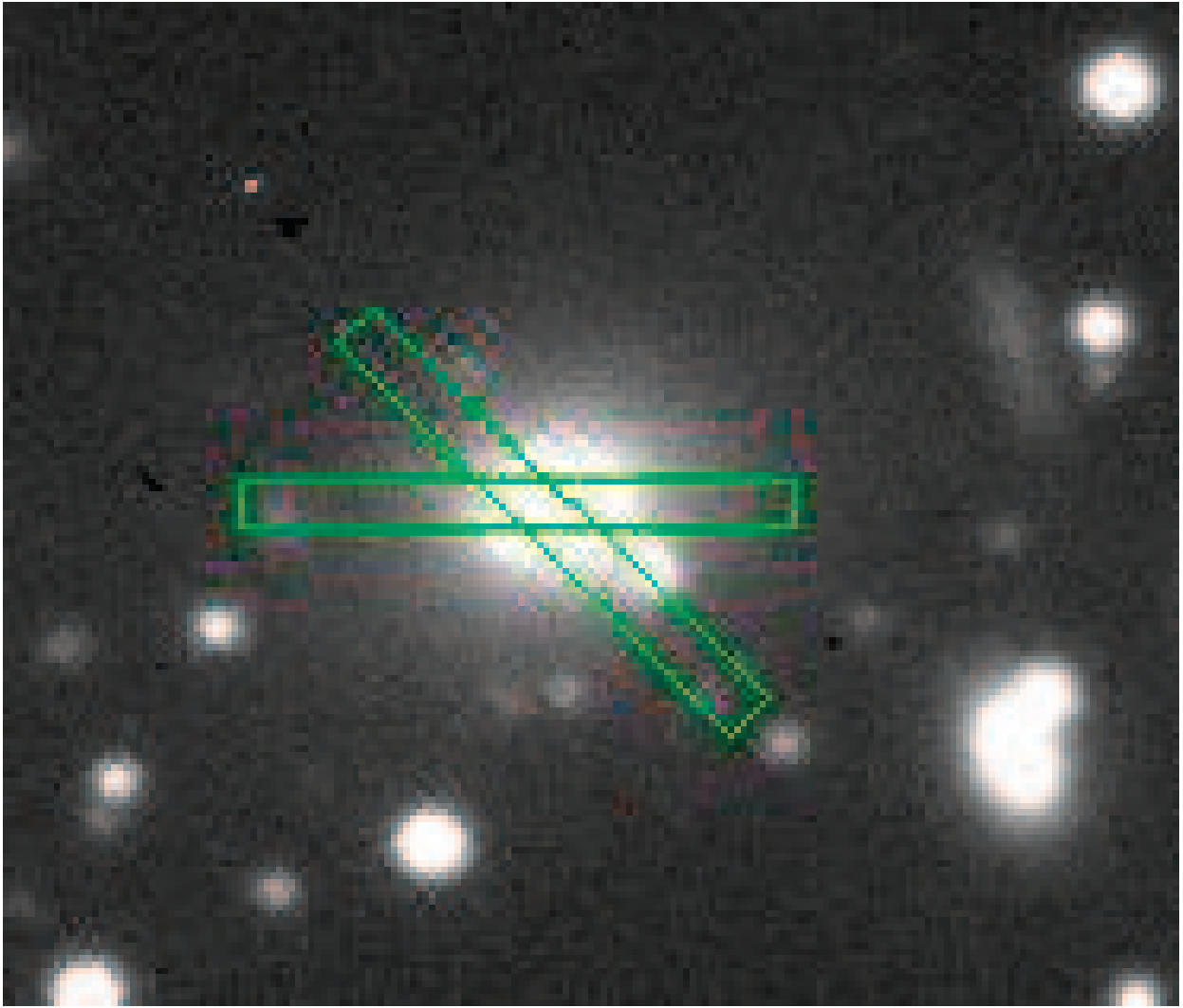}
\caption{VLT/FORS2 acquisition image of PKS0745-19 prior to spectroscopic observations, oriented North-South, with long slit locations illustrated.\label{fig3}}
\end{figure}

\begin{figure}
\includegraphics[width=6.5in]{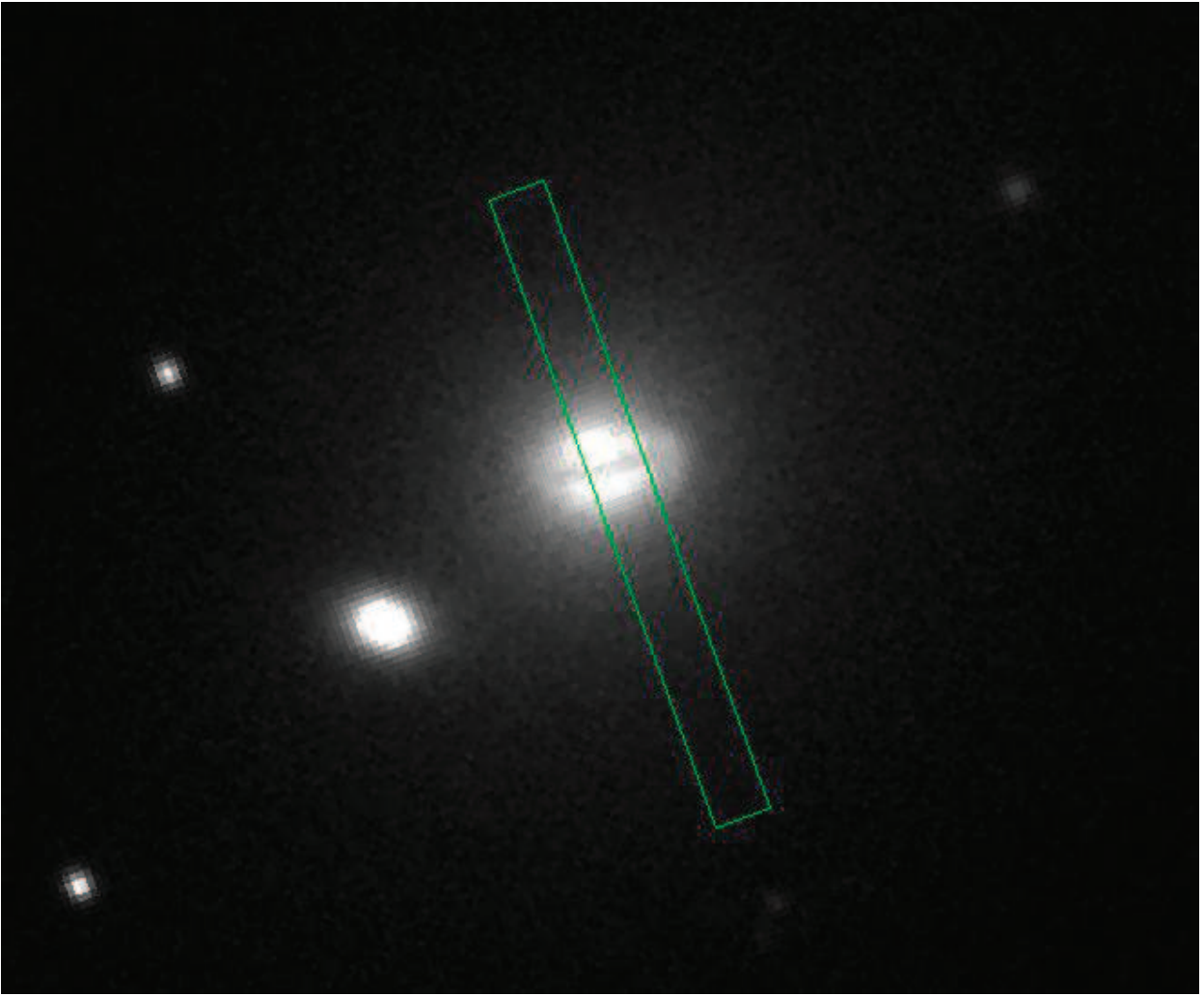}
\caption{VLT/FORS2 acquisition image of Hydra A prior to spectroscopic observations, oriented North-South, with long slit location illustrated.\label{fig4}}
\end{figure}

\begin{figure}
\includegraphics[width=6.5in]{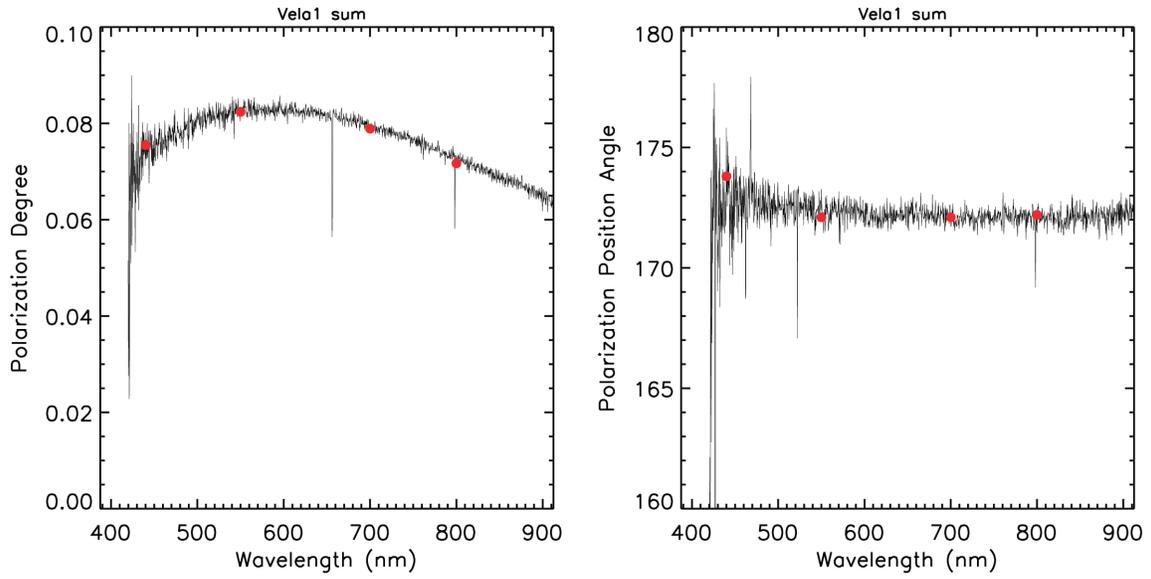}
\caption{Observations of polarized standard star. Black lines are data, and filled circles (red in electronic version) show expected value from literature, see also Table~2.\label{fig5}}
\end{figure}

\begin{figure}
\includegraphics[width=6.5in]{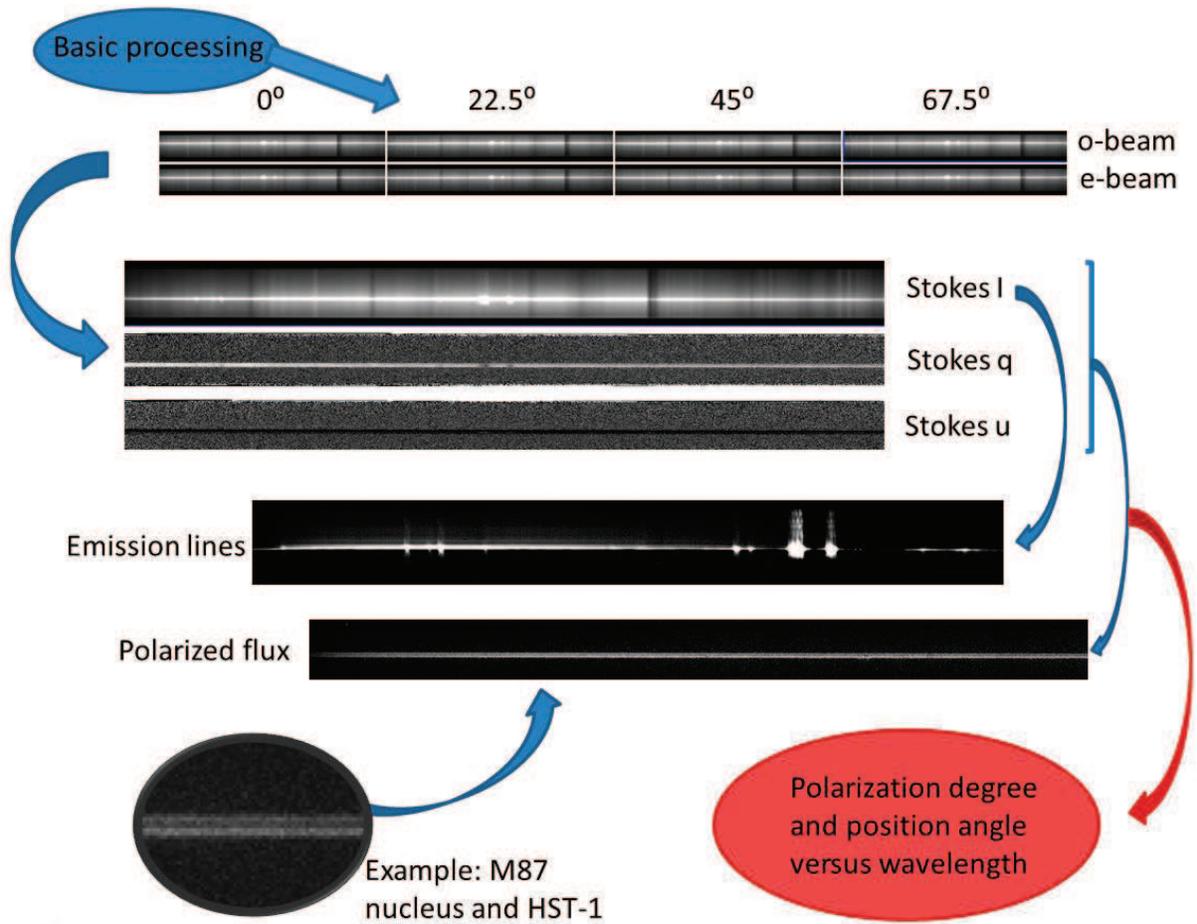}
\caption{Illustration of the data processing stages required to derive
  polarization information using FORS longslit spectropolarimetry
  mode.The o- and e-beam images are derived from basic image
  processing as described in the text, and these are combined in the
  appropriate way to yield polarization images.\label{fig6}}
\end{figure}

\begin{figure}
\includegraphics[width=6.5in]{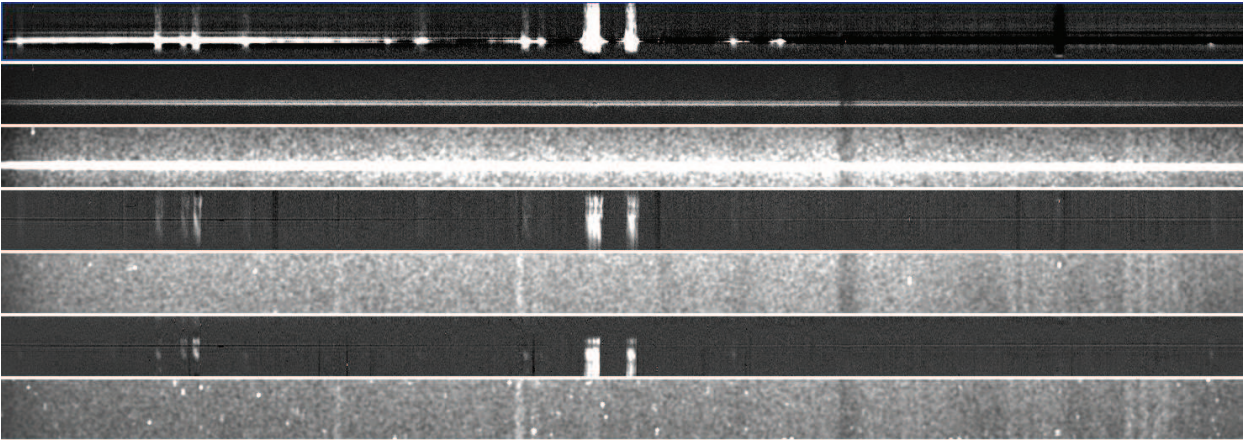}
\caption{Long slit spectropolarimetric data for M87. Panels from top
  to bottom, are (\romannumeral1) Slit 1 (nucleus) line emission
  image, followed by (\romannumeral2) the polarized flux image at low
  contrast (\romannumeral3) the polarized flux image at high contrast
  (\romannumeral4\ \& \romannumeral5) Slit 2 line emission and
  polarized flux (\romannumeral6\ \& \romannumeral7) Slit 3 line
  emission and polarized flux.\label{fig7}}
\end{figure}

\begin{figure}
\includegraphics[width=6.5in]{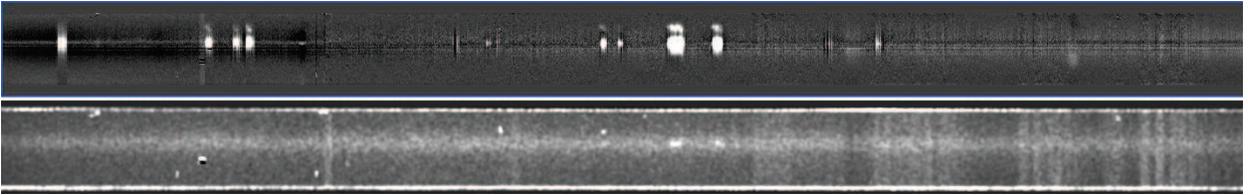}
\caption{Long slit spectropolarimetric data for Hydra A. Panels from
  top to bottom, are line emission image, followed by the polarized
  flux image.\label{fig8}}
\end{figure}

\begin{figure}
\includegraphics[width=6.5in]{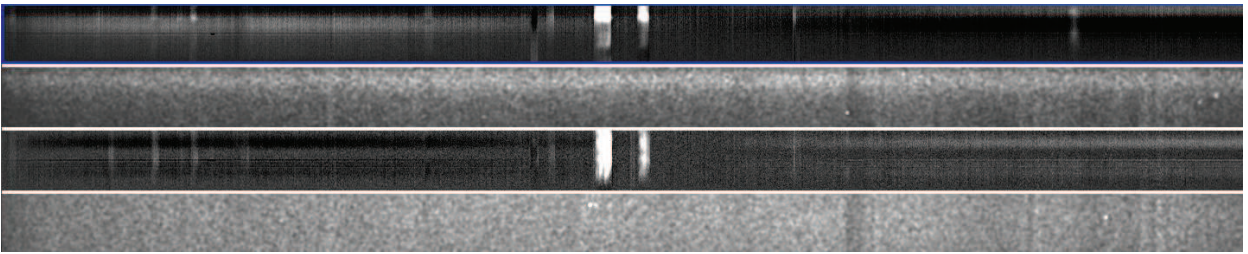}
\caption{Long slit spectropolarimetric data for NGC4696. Panels from
  top to bottom, are Slit 1 (nucleus) line emission image, followed by
  the polarized flux image; Slit 2 (dust lane) line emission image and
  polarized flux image.\label{fig9}}
\end{figure}

\begin{figure}
\includegraphics[width=6.5in]{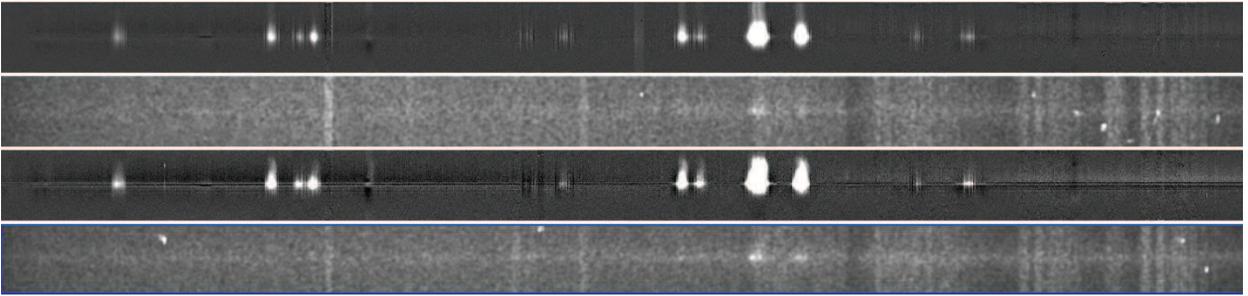}
\caption{Long slit spectropolarimetric data for PKS0745-19. Panels
  from top to bottom, are slit p.a. 90$^{\circ}$ line emission image,
  its polarized flux image; slit p.a. 45$^{\circ}$ line emission image
  and polarized flux image.\label{fig10}}
\end{figure}

\begin{figure}
\includegraphics[width=6.5in]{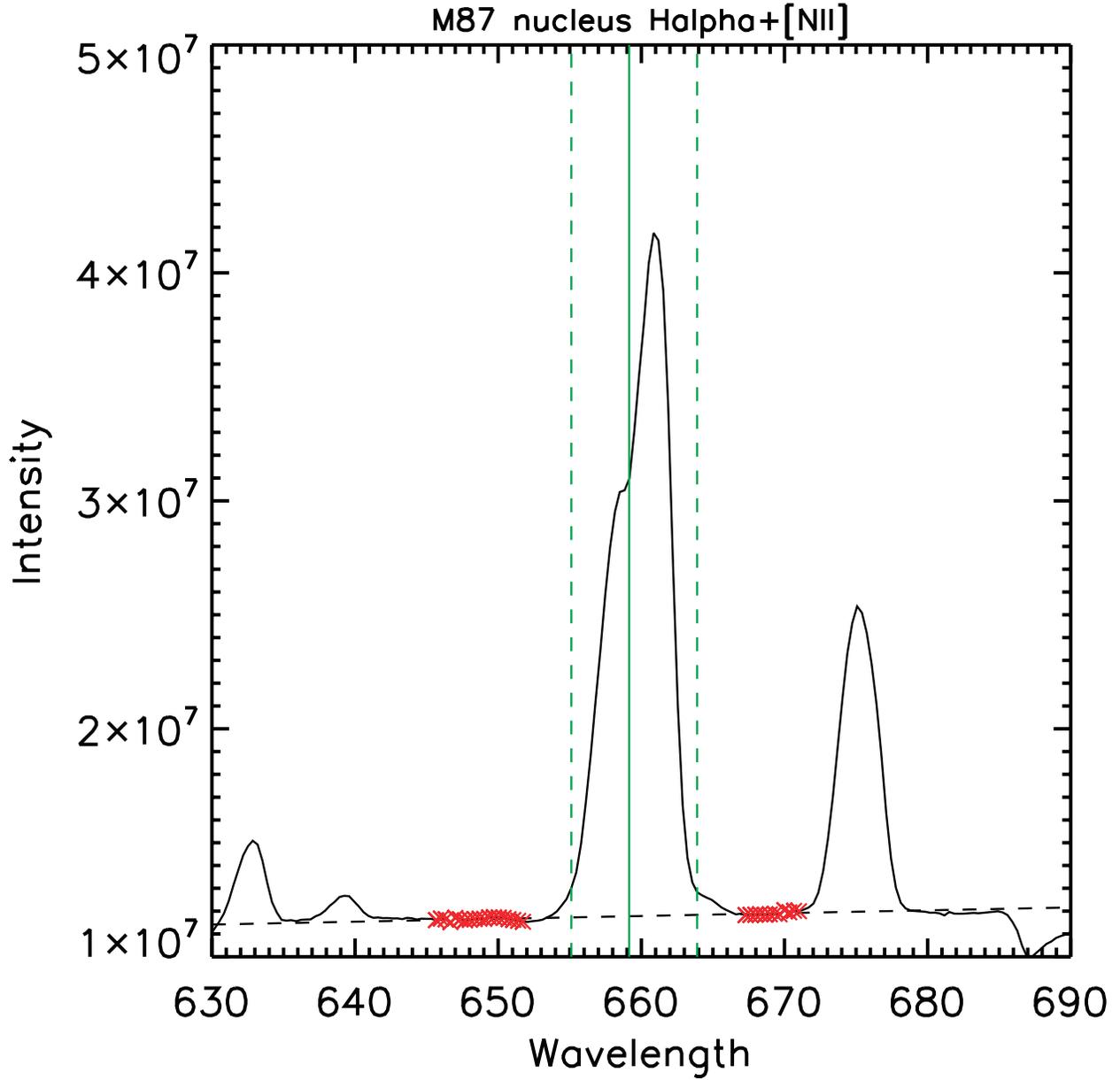}
\caption{Extracted line spectrum of the $H\alpha+[NII]$ region of the M87 nucleus. The vertical lines show the fiducial wavelength and the edges of the line-emission region. The crosses (red in the electronic version) show the selected continuum points, and the dashed line the continuum fit.\label{fig11}}
\end{figure}

\begin{figure}
\includegraphics[width=6.5in]{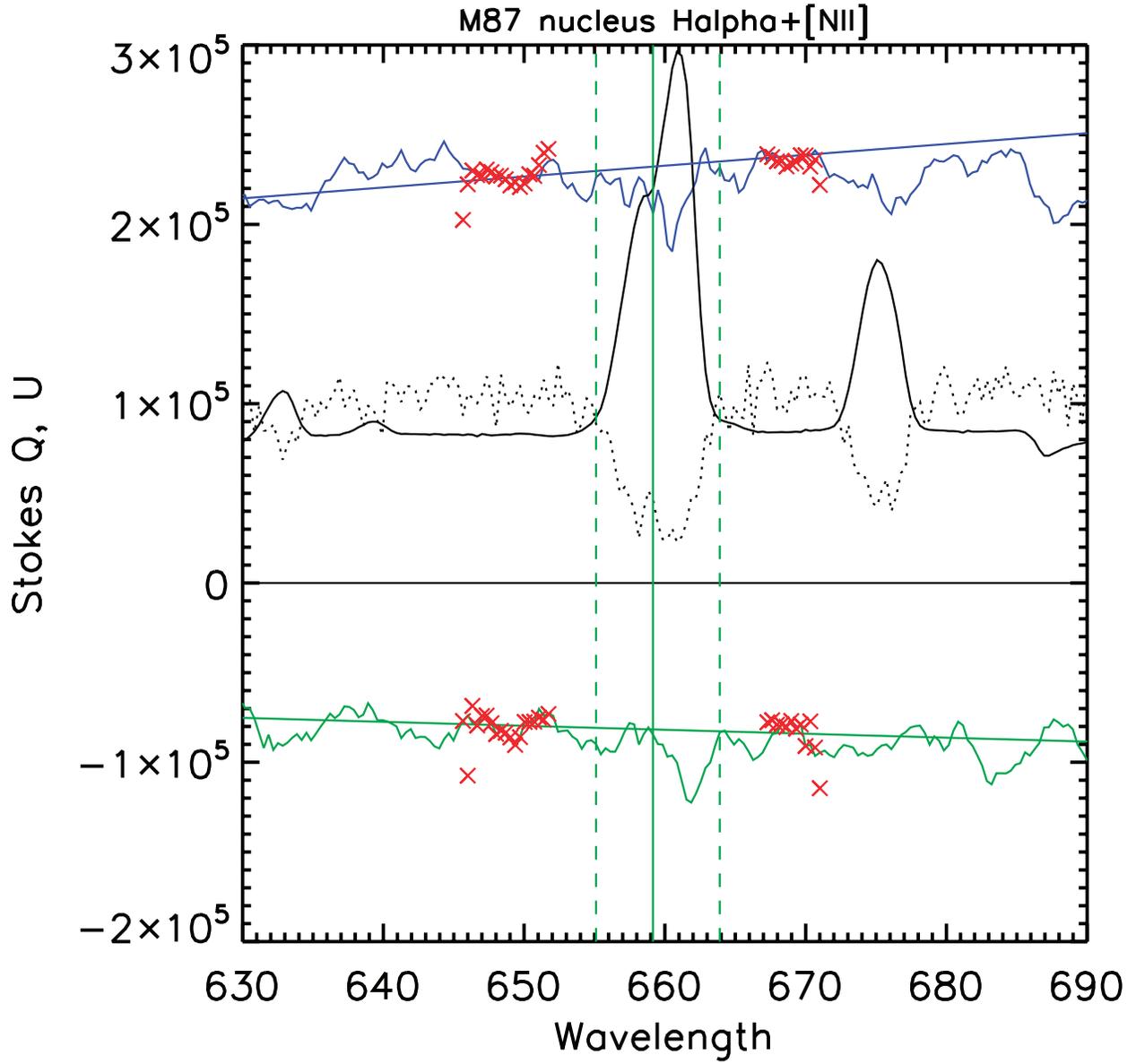}
\caption{Stokes parameters through the same regions indicated in Fig.~11, showing the expected polarized continuum emission. Stokes U is the lower (green) curve, Stokes Q is the upper (blue) with their individual continuum fits shown as straight lines fitted through the region with the (red) crosses. The solid black line shows Stokes I scaled arbitrarily to fit, and the dotted black line shows the polarization degree multiplied by $5\times 10^6$, for comparison.\label{fig12}}
\end{figure}

\begin{figure}
\includegraphics[width=6.5in]{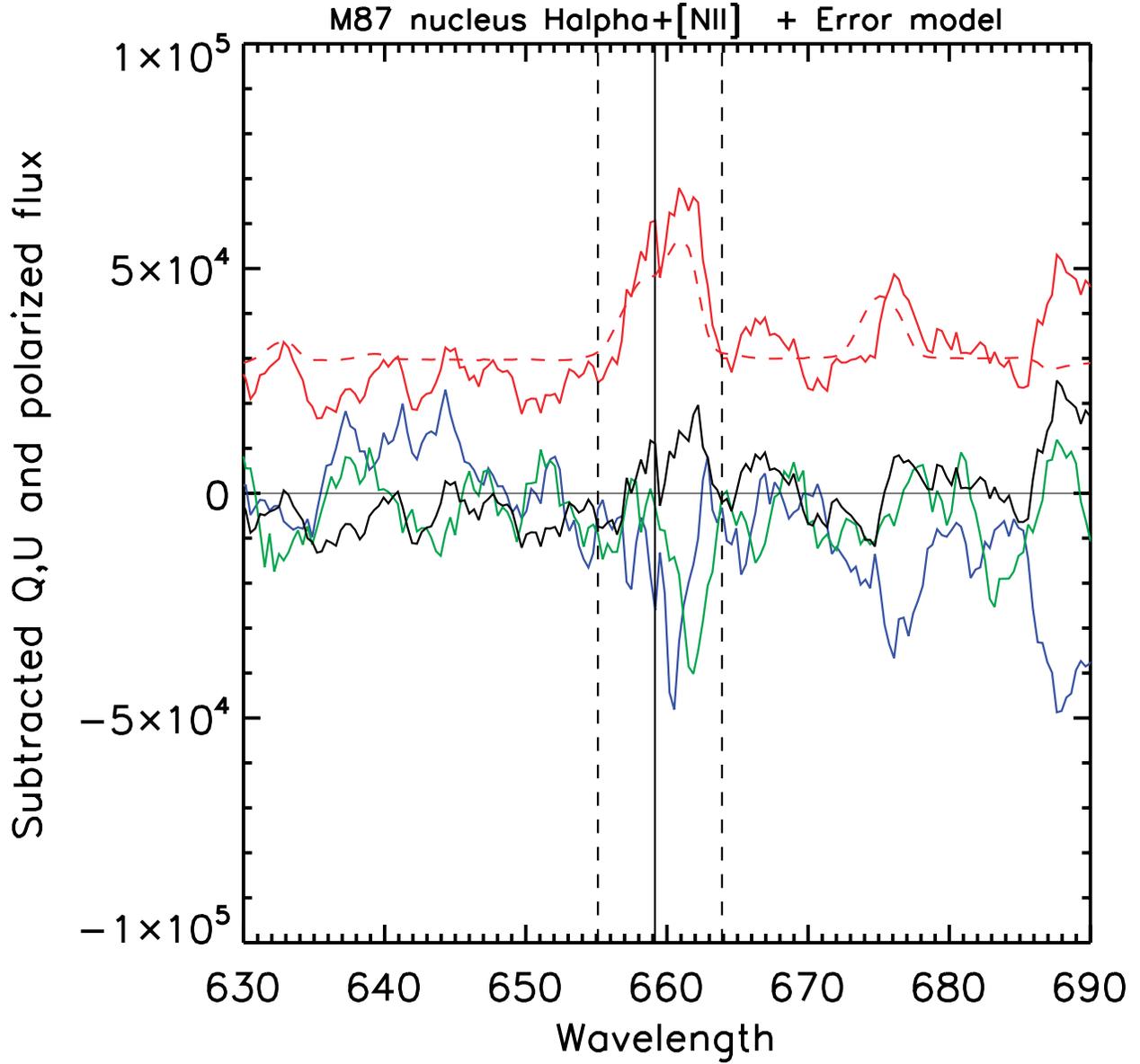}
\caption{Continuum subtracted Stokes Q (blue), Stokes U (green) are shown with their implied polarized flux (red). The Poisson noise model for polarized flux assuming no intrinsic polarization is shown as dashed red. The noise corrected `debiassed' polarized flux is plotted as the central black line, which remains close to zero through the emission region.\label{fig13}}
\end{figure}

\begin{figure}
\includegraphics[width=6.5in]{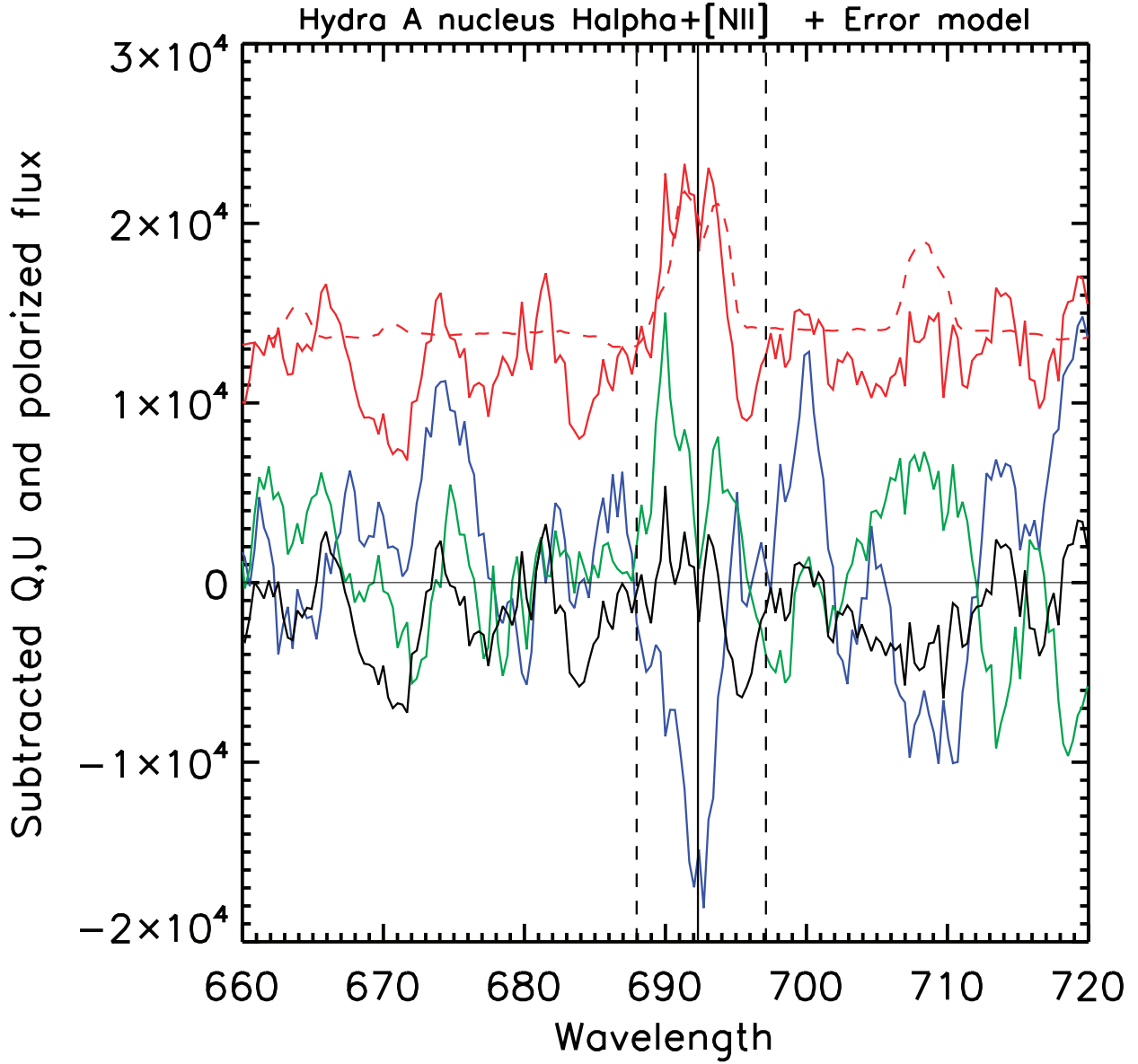}
\caption{Continuum subtracted Stokes Q (blue), Stokes U (green) and polarized flux(red) with polarized flux noise model (dashed red) for Hydra A. The central black line is the noise corrected debiassed polarized flux. On correction for the noise model, the apparent excess of polarized emission at the location of the emission lines
is seen to be a consequence of the positive definite nature of the polarized flux.
The final value for the polarization degree of the line complex, $\approx 0.1\% $, is not significantly above the uncertainty.
\label{fig14}}
\end{figure}


\end{document}